\documentclass[%
 reprint,
longbibliography,
 amsmath,amssymb,
 aps,
prb,
]{revtex4-1}
\setcitestyle{numbers,square}

\usepackage{braket}
\usepackage{graphicx}
\usepackage{dcolumn}
\usepackage{bm}
\usepackage[normalem]{ulem} 
\usepackage{color}
\usepackage{siunitx}
\usepackage[version=4]{mhchem}

\DeclareSIUnit\elementarycharge{\text{\ensuremath{e}}}
\DeclareSIUnit\angstrom{\text {Å}}
\DeclareSIUnit\uc{\text {unit-cell}}

\usepackage{hyperref}
\hypersetup{
    colorlinks,%
    citecolor=blue,%
    linkcolor=blue,%
    urlcolor=blue
}

\newcommand{\orcid}[1]{\href{https://orcid.org/#1}{\includegraphics[width=8pt]{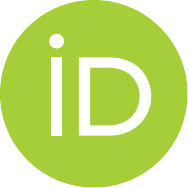}}}

\begin{document}


\title{Vacancy localization effects on MX$_2$ transition metal dichalcogenides: a systematic ab-initio study}

\author{Rafael L. H. Freire\orcid{0000-0002-4738-3120}}
\email{rafael.freire@lnnano.cnpem.br}

\author{Felipe Crasto de Lima\orcid{0000-0002-2937-2620}} 
\email{felipe.lima@ilum.cnpem.br}

\author{Adalberto Fazzio\orcid{0000-0001-5384-7676}}
\email{adalberto.fazzio@ilum.cnpem.br}

\affiliation{Ilum School of Science, CNPEM, 13083-970 Campinas, Brazil}

\date{\today}

\begin{abstract}
Two-dimensional transition metal dichalcogenides (MX$_2$) vacancy formation energetics is extensively investigated. Within an ab-initio approach we study the MX$_2$ systems, with M=Mo, W, Ni, Pd and Pt, and X=S, Se, and Te. Here we classify that chalcogen vacancies are always energetic favorable over the transition metal ones. However, for late transition metals Pd $4d$, and Pt $5d$ the metal vacancy are experimentally achievable, bringing up localized magnetic moments within the semiconducting matrix. By quantifying the localization of the chalcogen vacancy states we evidentiate that it rules the intra- and inter-vacancy interactions that establish both the number of vacancy states neatly lying within the semiconducting gap, as well as its electronic dispersion and SOC splitting. Combining different vacancies and phase variability 1T and 1H of the explored systems allow us to construct a guiding picture for the vacancy states localization.

\end{abstract}

\maketitle

\section{Introduction}

Defects on materials, and particularly atomic vacancies, arises as a possible route for an experimental control of properties \cite{ACSNANOliang2021, CPCcavallini2022}. Although the plethora of possible materials defects, ranging from localized to extended, even the elementary structural defect (atomic single vacancy) allow for an complex phenomena to arise, for instance ruling the molecular self-assemblies over two-dimensional (2D) substrates \cite{NATMATlin2017}. Additionally, controlling such punctual defects on the precise manipulation of vacancies have allowed the construction and engineering of metamaterials \cite{PRLnguyen2018, ACSNANOli2015}. In order to further explore materials modification through vacancies, a clear picture of the intrinsic vacancy properties are needed. 

Two-dimensional transition metal dichalcogenides (TMD's) are an interesting  class of materials holding tremendous potential for applications, encompassing catalysis \cite{ADVMATvoiry2016, NATCOMtsai2017}, (opto)electronics \cite{NLcheng2014, NATNANOcui2015, NATNANOwang2012}, spintronics \cite{NATNANOwang2012, RMPavsar2020}, magnetism \cite{NATCOMavsar2020} and energy storage \cite{CCdu2010, ACSNchang2011}. Particularly, the presence of defects in 2D-TMD materials can directly impact their properties, causing a variety of phenomena that could be either detrimental acting like carrier scattering and localization sites \cite{NLedelberg2019} or beneficial as is the case of active catalytic sites, and magnetic orderings \cite{JACScai2015}. Furthermore, spin-orbit effects on those materials combined with point defects may give rise to interesting phenomena such as giant magnetoresistance \cite{JPCMbai2018}, topological phases \cite{NLcrasto2021}, all of which could be employed on spintronic devices. Additionally, recent experimental works have shown that the localization character of the vacancy states induces electronic transport \cite{NATCOMqiu2013}, and magnetism ordering on transition metal vacancies \cite{PRBabsor2017}.

\begin{figure}[h!]
\includegraphics[width=\columnwidth]{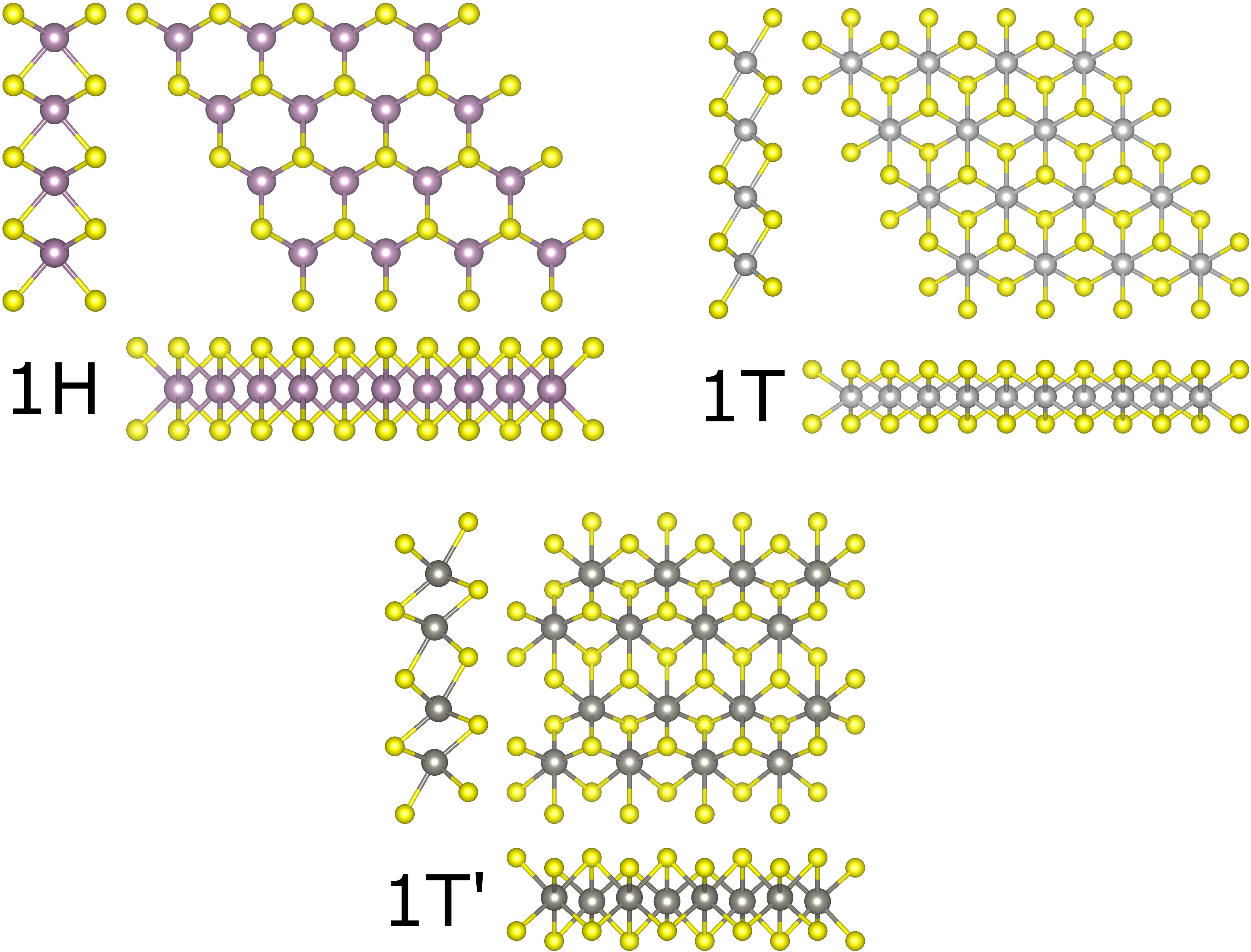}
\caption{1H, 1T, and 1T' MX$_2$ 2D structural phases.}
\label{fig:ucell}
\end{figure}

In this study, we perform ab-initio calculations based on the density functional theory to clarify and quantify the localization character of vacancies states on TMD. We considered the pristine and defected MX$_2$ 2D systems (for M $=$ Ni, Mo, Pd, W, and Pt, and X $=$ S, Se, and Te) in their 1H, 1T, and 1T' phases (see Fig.~\ref{fig:ucell}). We highlight the most stable structural phase and most stable vacancies (of metals or chalcogen). We present an analysis of the localization nature of such vacancy states and their consequences to the energetic, electronic, and magnetic properties.

\section{Computational Approach}

Spin-polarized calculations based on density functional theory (DFT) \cite{PRhohenberg1964, PRkohn1965} were performed within the semi-local exchange-correlation functional proposed by Perdew--Burke--Ernzerhof (PBE) \cite{PRLperdew1996}. To treat the long range dispersion van der Waals (vdW) interactions, the pairwise D3 correction framework as proposed by Grimme were considered  \cite{JCPgrimme2010, JCCgrimme2011}.

\begin{figure*}[ht!]
\includegraphics[width=2\columnwidth]{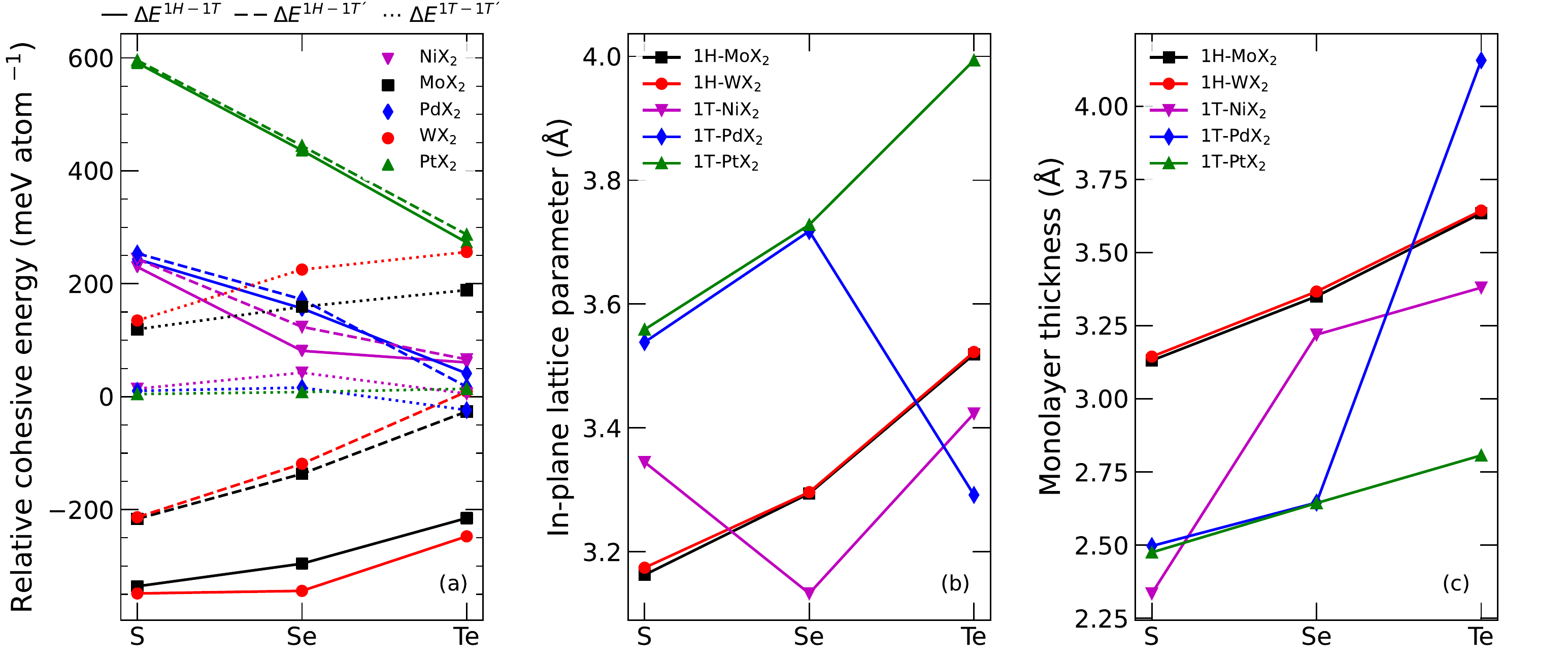}
\caption{(a) Relative cohesive energies for all pristine systems and structural parameters, namely (b) in-plane lattice and (c) monolayer thickness, for the most stable MX$_2$ structures.}
\label{fig:cohen}
\end{figure*}

For the total energies, $E_{tot}^{\text{DFT+vdW}}$, the electron-ion core interactions were considered within the the projector augmented wave (PAW) method \cite{PRBblochl1994, PRBkresse1999}, as implemented in the  Vienna {\it Ab-Initio} Simulation Package (VASP) \cite{PRBkresse1993, PRBkresse1996}. The spin-orbit coupling was considered for the most stable configuration. For all calculations, cutoff energy for the plane-wave expansion of the Kohn--Sham orbitals was set to $520$\,eV, under an energy convergence parameter of $10^{-6}$\,eV, with all atoms relaxed until the atomic forces on every atom were smaller than $10^{-3}$\,eV\,{\AA}$^{-1}$. A uniform $4\times4\times1$ k-point mesh was considered for the Brillouin zone (BZ) integration, while a thinner mesh grid, $8\times8\times1$, was used for Density of States (DOS) calculations.

The 2D in-plane lattice parameters ($a$ and $b$) were optimized for a $1\times1$ unit cell with a fixed vacuum distance ($c$) of at least $20$\,{\AA}. Those lattice parameters were used to build up the pristine and single vacancy $5\times5$ and $3\times3$ unit cells to the 1H and 1T (1T') structures, with the former being large enough to avoid spurious interaction between periodic images.

\section{Results and discussion}

\subsection{Structural Stability}

To analyze the structural stability of the pristine structures, we calculated the cohesive energies -- being the total energy difference between the material and each isolated atom -- for the 1H, 1T, and, 1T' phases. To better present the relative stability of one phase against each other, we evaluated their relative cohesive energy, namely $\Delta E^{1H-1T}$, $\Delta E^{1H-1T'}$, and $\Delta E^{1T-1T'}$, as shown in Fig.~\ref{fig:cohen}~(a). That is, $\Delta E^{1H-1T}>0$ indicates the 1H phase is more stable than the 1T phase, and similarly to the other relative energies. As one can see, the negative values for 1H-1T ($200$--$400$\,meV\,atom$^{-1}$) and 1H-1T´ ($0$--$200$\,meV\,atom$^{-1}$) for \ce{MoX2} and \ce{WX2} indicate the 1H phase as the most stable, with this stability decreasing with the increase of the chalcogen atomic radius. Despite the slightly positive values observed for \ce{NiX2}, \ce{PdX2}, and \ce{PtX2}, the cohesive energy differences point to the 1T phase as the most stable, with the 1T-1T' differences being close to zero. Although the calculation started in the 1T' phase, after relaxation it converged barrierlessly into the 1T phase.

In general, the 1T phase has larger in-plane lattice parameters [Fig.~\ref{fig:cohen}(b)] with a smaller monolayer thickness as compared to the 1H phase and its more open structure (hexagonal holes). Consequently the monolayer thickness [Fig.~\ref{fig:cohen}(c)] decreases as the chalcogen planes move down to stabilize the equilibrium covalent bond length. In the 1H phase, the in-plane lattice parameter and monolayer thickness increase going from S$\rightarrow$Te, as a result of accommodating larger chalcogen atoms within the structure. In the case of the 1T phase, only PtX$_2$ follows a similar trend as the 1H systems. On the other hand, 1T-NiX$_2$ and 1T-PdX$_2$ undergo larger relaxation effects.

\subsection{Vacancy formation energy}

After pristine systems optimization, we built up a $5\times5$ defective monolayers considering the most stable phases for each system, namely the 1H phase for MoX$_2$ and WX$_2$ systems, and the 1T phase for NiX$_2$, PdX$_2$, and PtX$_2$. Both native point defects comprising chalcogen (V$_X$) and transition metal (V$_M$) vacancies were considered with the vacancy-vacancy distances in the 1H (1T) around $15.8$--$17.6${\AA} ($15.7$--$20.0${\AA}).

\begin{figure}
\includegraphics[width=\columnwidth]{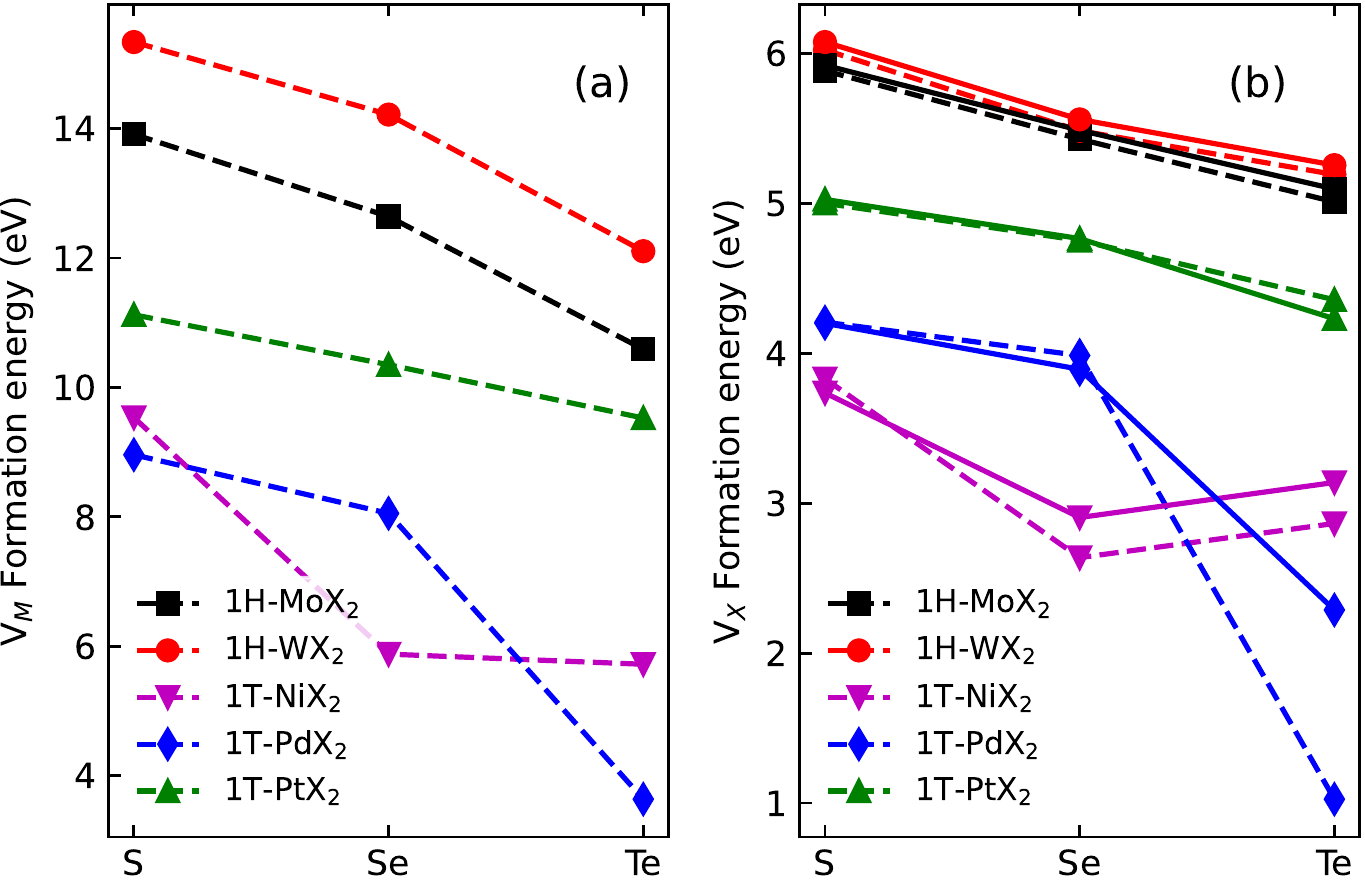}
\caption{Native point defect formation energies. (a) is the transition metal vacancy defect, and (b) the chalcogen defect. In the latter, we show formation energies for both $5\times5$ (dashed lines) and $3\times3$ (solid lines) supercell sizes.}
\label{fig:formen}
\end{figure}

In Figure~\ref{fig:formen}, we present the formation energies, $E_f$, for both M and X vacancies for the selected systems, evaluated according to
\begin{equation}
    E_f = E_{def}^{MX_2} - (E_{pristine}^{MX_2} - E_{M,X}^{free-atom} ),
    \label{eq:formen}
\end{equation}
in which $E_{def}^{MX_2}$ is the energy of the defected system; $E_{pristine}^{MX_2}$ is the energy of the pristine system; and $E_{M,X}^{free-atom}$ is free-atom energy of the corresponding atom M or X which generates the vacancy defect.

As indicated in Fig.~\ref{fig:formen} both formation energies are endothermic, with the formation energies of transition metal vacancies [Fig.~\ref{fig:formen} (a)] are higher than that for the chalcogen [Fig.~\ref{fig:formen} (b)]. Considering the different structure phases, one can realize that the defect formation energies in the 1T phase are always smaller than those for the 1H phase. Particularly, Ni and Pd systems with the heavier chalcogen (Se and Te) present transition metal vacancy formation energy close to the scale values of the chalcogen vacancies, indicating a possible occurrence of such defects (although less favorable than X-vacancies). Additionally, for the X-vacancies, we present the formation energy in a $3\times3$ supercell, i.e. increased vacancy density, where the interaction between neighboring vacancies is greater. As we can see the NiSe$_2$, NiTe$_2$, and PdTe$_2$ structures present a larger difference in the chalcogen vacancy formation energy in the $3\times3$ cell when compared with the lower vacancy density case. As we will show in the next sections, such behavior is a consequence of the more delocalized nature of the vacancy states in those systems.

\subsection{Band structure analysis}

\begin{figure*}
\includegraphics[width=2\columnwidth]{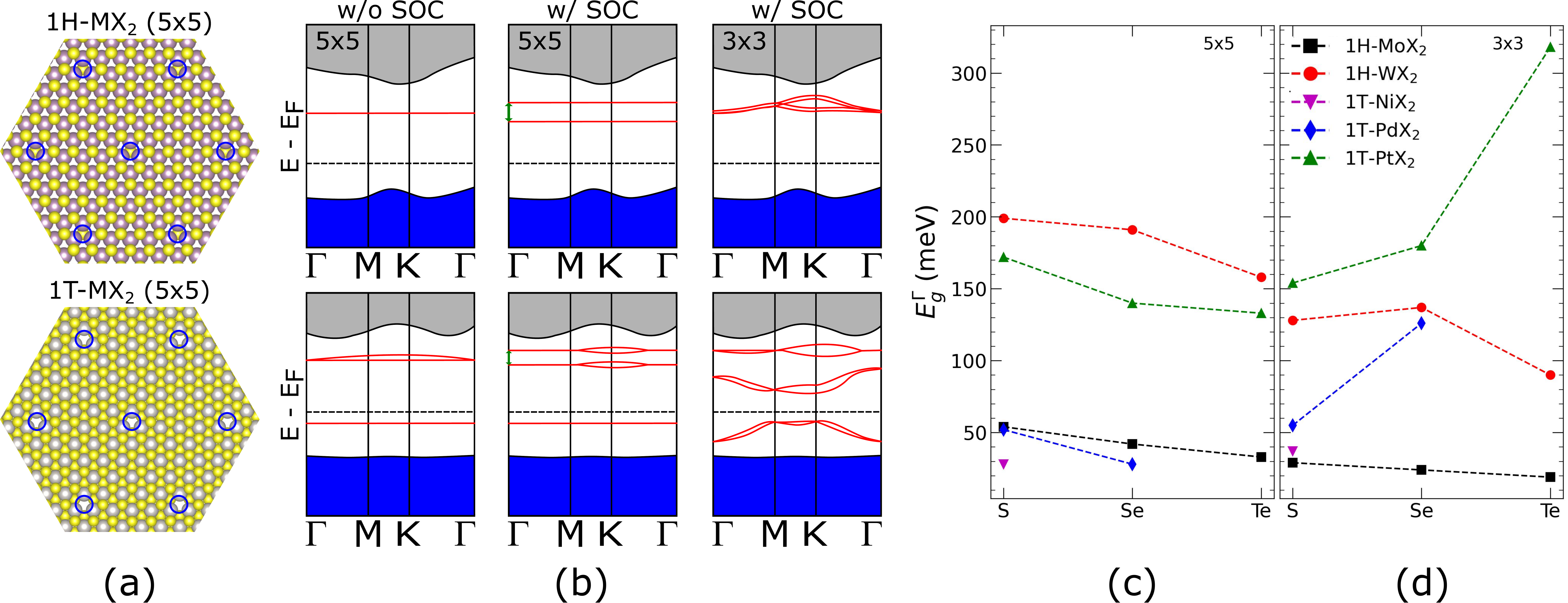}
\caption{(a) $5\times5$ defect supercell for 1H and 1T phases as indicated; (b), the universal schematic of band structure of each semiconducting phase, respectively, with and without spin-orbit corrections. The green arrows indicated spin-orbit splitting of the corresponding $k$-point; (c) Spin-orbit splittings calculated for both $3\times3$ and $5\times5$ defect supercells. $E^{\Gamma}_g$ is the spin-orbit splitting at the $\Gamma$ point.}
\label{fig:soc-bands}
\end{figure*}

In Fig.~\ref{fig:soc-bands}, we show a defected $5\times5$ supercell model for both 1H and 1T structures alongside their characteristic band structures with and without spin-orbit coupling. The introduction of the chalcogen vacancy leads to three defect states, corresponding to the three transition metal dangling bonds over the vacancy environment. The triangular $C_{3v}$ local environment split the dangling bond states into $E$ and $A_1$ irreducible representations. Additionally, the interaction of those three intra-vacancy dangling bonds in the 1H structure is larger than the 1T, given the remaining chalcogen coupling those TM. That is, this stronger coupling leads to a large energy separation between the $E$ and $A_1$ irreducible representation. Such an interaction picture leads to three vacancy states on the 1T phase neatly lying on the system energy gap ($E$ and $A$ states), while for the 1H phase the two $E$ states remain in the energy gap, with the $A_1$ state being within the valence bands. Additionally to this picture, the SOC effect has a role in splitting the $E$ states into two sets. Such value of splitting is directly related to the SOC contribution in each system. In Fig.~\ref{fig:soc-bands}(c) we show the SOC splitting value at $\Gamma$ for the lower vacancy density system ($5\times5$ cell). Here, the TM atom mostly rules the SOC contribution, with a lower variance of the value with the chalcogen atom.  Such variance is higher for WX$_2$ systems, ~50\,meV, while lower for MoX$_2$ systems, ~25\,meV. For the 5d TM, the 1H tungsten phases present the highest splitting, with the WS$_2$ reaching close to 0.2\,eV, followed by the 1T Pt phases. For the 4d TM similar trend is observed with the 1H Mo phases presenting higher SOC splitting than Pd 1T phases. Interestingly, going towards a higher vacancy density, where the interaction between adjacent vacancy states becomes important, the SOC trend is altered. First note that the vacancy states become dispersive, Fig.~\ref{fig:soc-bands}(b) right panels. The combined effect of the dispersion of the bands and SOC, induce a higher gap opening at the $\Gamma$ point for the 1T phases compared with the 1H ones. For instance, such a gap goes over 0.3\,eV for PtTe$_2$, with a dependence of the chalcogen atom being more accentuated. The dependence arises given the indirect coupling between adjacent vacancy states being ruled by the TMD matrix \cite{NLcrasto2021}. To further explore such indirect coupling we quantify the localization of such vacancy states.

\subsection{(De)Localization}

\begin{figure}
\includegraphics[width=\columnwidth]{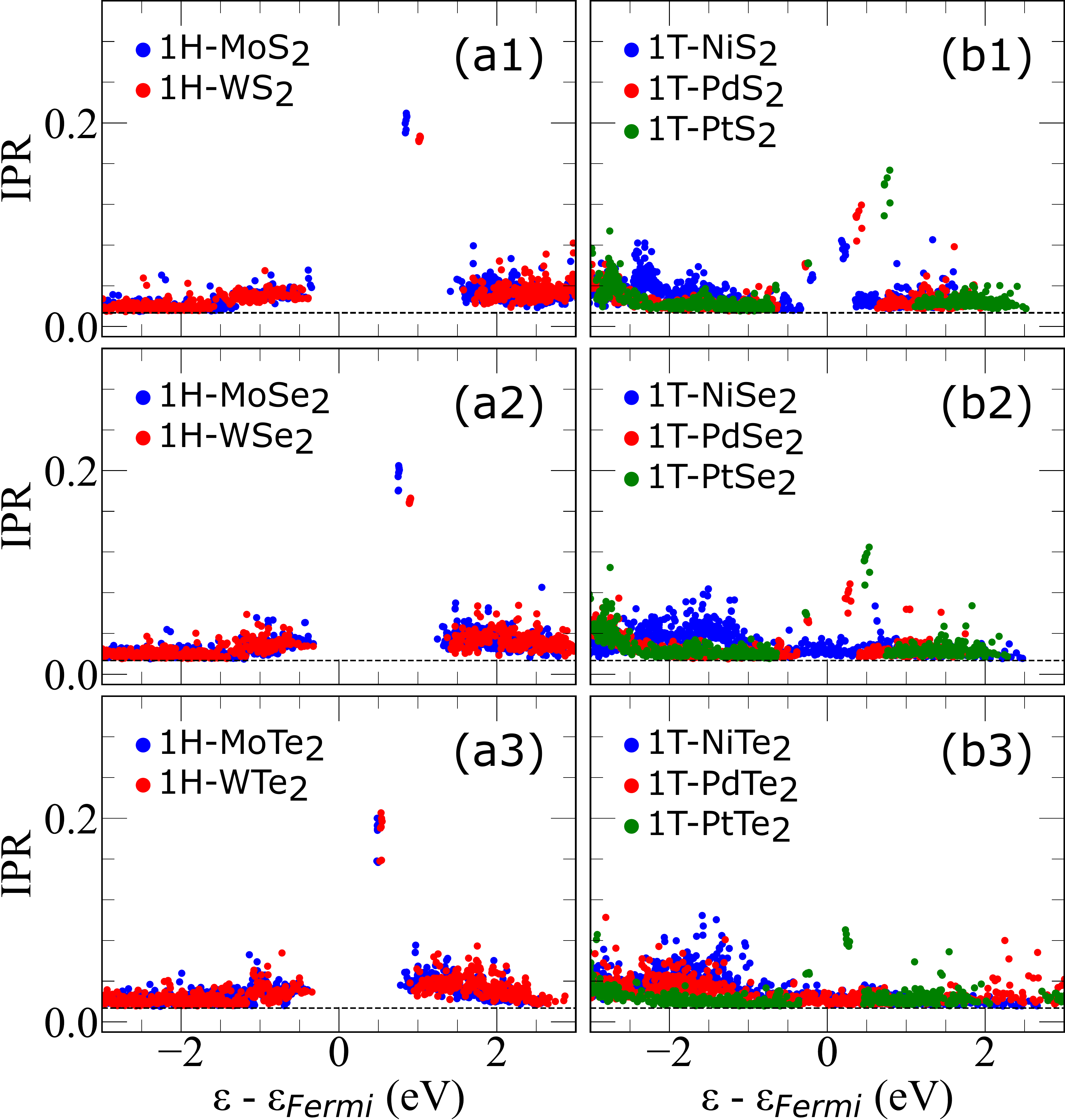}
\caption{Inverse participation ratio (IPR) as a function of the Bloch wavefunction ($|\psi_{n,k}\rangle$) energy. (a1)-(a3) The IPR for 1H-MX$_2$ phases with M$=$Mo, and W and (b1)-(b3) The IPR for 1T-MX$_2$ phases, with M$=$Ni, Pd, and Pt from S to Te, respectively}
\label{fig:ipr}
\end{figure}

To take into account (de)localization effects due to the vacancy formation, we characterized the defective systems through the inverse participation ratio (IPR) \cite{PRBfocassio2021} given by
\begin{equation}
 \text{IPR}_{n,k} = \frac{ \sum\limits_{i=1}^{N} | \langle i | \psi_{n,k} \rangle |^4}{ \left( \sum\limits_{i=1}^{N} | \langle i | \psi_{n,k} \rangle |^2 \right)^2},
\label{eq:ipr}   
\end{equation}
in which $\langle i|\psi_{n,k}\rangle$ was taken as the sum of the orbital projected KS eigenstate for each site/atom $i$, such that $N$ is the total number of atoms in the cell. Thus, for a fully localized state the IPR should be one, while a delocalized one corresponds to the limit $\text{IPR} = 1/N$. In Fig.~\ref{fig:ipr}, we show the IPR values for each $| \psi_{n,k} \rangle$ state as a function of its eigenvalue, for the most stable phases for chalcogen vacancies. It is worth pointing out that those defective systems are in general semiconducting, with exception of NiSe$_2$, NiTe$_2$, and PdTe$_2$, which are metallic, as indicated by the energy gap at Fig.~\ref{fig:ipr} $x$-axis.

\begin{figure}
\includegraphics[width=\columnwidth]{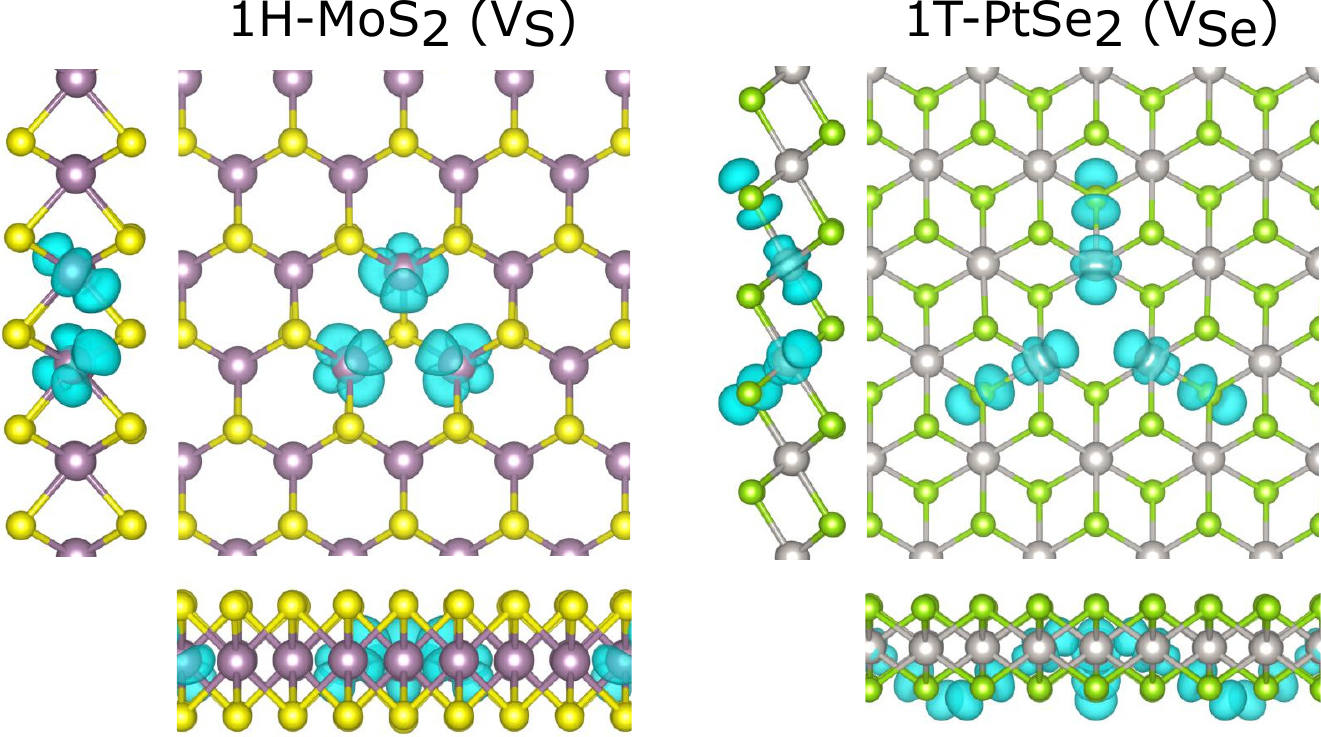}
\caption{LDOS for the chalcogen-defected systems (a) 1H-MoS$_2$ and (b) 1T-PtSe$_2$, with the same isosurface of 0.007\,e/{\AA}$^3$}.
\label{fig:ldos}
\end{figure}

In the semiconducting cases, the localized states that appear within the gap corresponding to the vacancy states as depicted in Fig.\,\ref{fig:soc-bands}. The 1H phases present a slightly larger IPR when compared to the 1T phases. Taking the 1H phase as an example [Fig.~\ref{fig:ipr}(a1)-(a3)], after the chalcogen vacancy formation, if (in the ideal case) only the three transition metal orbitals ($|M_i \rangle$) neighboring the vacancy contributes equally (with $\langle M_i |  \psi_{vac-state}\rangle =a$) to the localization, we have $\text{IPR} = 3a^4 / (3a^2)^2 = 1/3$. That is, the localization limit is $1/3$ at the chalcogen vacancy surroundings. As observed the IPR values are around $0.20$ which is close to the ideal limit, being reduced due to the spread of the vacancy states to hybridized neighboring orbitals. The same IPR limit (IPR$=1/3$) is valid for the 1T structures [Fig.~\ref{fig:ipr}(b1)-(b3)]. However, despite the vacancy states being within the bandgap, the values observed are slightly small, lying around IPR$=0.15$. In this case, an enhanced environment interaction is observed. To better visualize it, in Fig.~\ref{fig:ldos} we show the vacancy states squared wave function (partial charge density) of MoS$_2$ and PtSe$_2$ with S and Se vacancy, which respectively present an IPR of $0.2$ and $0.1$. The 1T phase forms a pyramidal-like configuration [see Fig.~\ref{fig:ldos} (b)], in which the localized state spreads to opposite surface chalcogen atoms neighboring the dangling bond Pt atom. On the other hand, the 1H phase LDOS [see Fig.~\ref{fig:ldos}(a)] are mostly localized in the M atoms close to the vacancy. Thus, the IPR for these localized states decreases for the 1T phase as compared to the 1H one. The spatial distribution of those vacancy states, although with different spreads follow the same three-fold symmetry of the vacancy structure.

In the light of the quantification of the localization through the IPR, the chalcogen vacancy formation energy and the band structure dependence of the vacancy density can be readily explained. For instance, given the more localized nature of Mo, W, and Pt vacancy states all above IPR$=0.1$, dictates that the adjacent vacancies will have lower interaction not changing the formation energy significantly. However Ni and Pd, with their IPR$<0.1$ indicate a stronger interaction between adjacent vacancies reducing the formation energy for higher vacancy densities. Particularly, for the 1T phases with Te [Fig.~\ref{fig:ipr}(b3)] which presents the lower IPR values for the vacancy states, presents also the greater variability on the chalcogen vacancy formation energy, see Fig.~\ref{fig:formen}(b). A similar analysis can be done following the band structure. Here, the 1T phases allowing a longer range of interaction between adjacent vacancies leads to higher dispersive states when compared with the 1H structure [Fig.~\ref{fig:soc-bands}(b)]. Here the localization of the chalcogen vacancy states leads to the interpretation of their interaction, additionally, TM vacancies can also lead to localized magnetic effects.

\subsection{Magnetism}

Although less energetically favorable, TM vacancies can also be found on TMD, where the introduction of such vacancies can induce a local net magnetic moment. In Table~\ref{tab:mag} we summarize the systems presenting a net magnetic moment after a TM vacancy is formed, while the ones not present on the table did not present any magnetic properties. After the TM vacancy is formed, localized magnetic moments arise on the neighboring chalcogen atoms. Here a ferromagnetic (FM) phase can be stabilized for some explored systems, being the only observed phase for the 1H structure. Interestingly, some of the 1T systems can present an antiferromagnetic (AFM) arrange of such chalcogen magnetic moments, which can be more stable than the ferromagnetic, as already observed for 1T-PtSe$_2$ \cite{NATCOMavsar2020}. Aware of this behavior, we probed this antiferromagnetic configuration for our 1T-based based systems.
The systems which presented a possible antiferromagnetic phase were 1T-NiS$_2$, 1T-PdS$_2$, 1T-PdSe$_2$, 1T-PtS$_2$, and 1T-PtSe$_2$. However, as shown in Table~\ref{tab:mag}, for Ni- and Pd-based systems the antiferromagnetic phase ($\Delta E^{AFM-FM}<0$) were not the most stable. On the other hand, for Pt-based systems we found more stable antiferromagnetic phases for 1T-PtS$_2$ and 1T-PtSe$_2$ systems with an energy difference about $-54$\,meV/vacancy, and $-33$\,meV/vacancy, respectively. That is, such values for the AFM phase together with the FM of 1T-PdSe$_2$ [with $\Delta E^{AFM-FM}=43$\,meV/vacancy], dictates such magnetic configurations to be robust close to the ambient temperature.

\begin{table}[h!]
\begin{ruledtabular}
\caption{Magnetic moments, $m$ ($\mu_B$), induced in the most stable phases with the introduction of vacancies. The last column, $\Delta E^{AFM-FM}$  (meV/vacancy) , is the energy difference between the antiferromagnetic (AFM) and ferromagnetic (FM) phases, with $\Delta E^{AFM-FM}<0$ indicating the AFM phase being more stable.}
\label{tab:lattparams}
\label{tab:mag}
\begin{tabular}{lccc}
MX$_2$   & vacancy &  $m$ & $\Delta E^{AFM-FM}$ \\
\hline
1H-MoSe$_2$ & Mo & 4.00 & -- \\
1H-MoTe$_2$ & Mo & 2.00 & -- \\
1H-WTe$_2$  & W  & 1.96 & -- \\
1T-NiS$_2$  & Ni & 4.00 & 19 \\ 
1T-PdS$_2$  & Pd & 4.00 & 17 \\ 
1T-PdSe$_2$ & Pd & 4.00 & 43 \\ 
1T-PtS$_2$  & Pt & 4.00 & -54 \\ 
1T-PtSe$_2$ & Pt & 4.00 & -33 
\end{tabular}
\end{ruledtabular}
\end{table}

\section{Conclusions}

We systematically investigated the energetic and electronic properties of a series of two-dimensional transition metal dichalcogenides (MX$_2$, with M$=$Ni, Mo, Pd, W, and Pt; and X$=$S, Se, and Te) presenting native point defects, namely chalcogen and transition metal vacancies, in different structural phases. Here, we found the chalcogen vacancy as the most stable for all systems, with lower formation energy in the 1T phase (Ni, Pd, and Pt systems) than those in the 1H phase (Mo and W systems). However, transition metal vacancies can still be found under experimental conditions. In this sense, our results show the appearance of localized magnetic moments induced by metal vacancies for 1T-PtS$_2$ and 1T-PtSe$_2$ that could be applied to design new 2D magnets. Furthermore, we have explored the localization effects of the chalcogen vacancy states. Such, localized states give rise to three energy levels that can be neatly lying on the TMD matrix energy gap. Here the localization strength, quantified by the defect states inverse participation rate, is shown to be greater in the 1H phases. This leads to both (i) a stronger repulsion between the three defect states (with $C_{3v}$, $E$ and $A_1$ irreducible representations) increasing the gap between $E$ and $A_1$ states, and (ii) giving rise to lower dispersion for higher vacancy densities, that is, present a lower vacancy-vacancy interaction. For the 1T phases, the more delocalized nature of the vacancy states gives rise to a stronger hopping-like interaction between adjacent vacancies. Additionally, we have shown that vacancy-vacancy interactions (given by their localization) ruled not only the materials band dispersion but also the SOC splittings. This investigation brings insightful discussions on the nature of energetic and electronic effects of vacancy defects within the different 2D-TMD material phases and vacancy concentration.

\section*{Acknowledgments}
The authors acknowledge financial support from the Brazilian agencies FAPESP (grant 20/14067-3 and 17/02317-2), INCT-Nanomateriais de Carbono, and Laborat\'{o}rio Nacional de Computa\c{c}\~{a}o Cient\'{i}fica for computer time.

\bibliography{bib}

\end{document}